\documentclass{article}

\usepackage{amsmath}
\usepackage{graphicx}
\usepackage[utf8]{inputenc}
\usepackage{subcaption}
\usepackage{tikz}

\usepackage{hyperref}
\usepackage{breakurl}

\usetikzlibrary{snakes}

\usepackage[normalem]{ulem}

\usepackage{hyperref}

\usepackage{colordvi}
\usepackage{xcolor}

\newcommand{\progname}[1]{\texttt{#1}}
\newcommand{\pkgname}[1]{\emph{#1}}
\newcommand{\codid}[1]{\href{https://www.crystallography.net/cod/#1.html}{\texttt{#1}}}

\title{Determination of bonding radii from small-molecule crystal structures}

\author{
  Eglė~Šidlauskaitė \and
  Andrius~Merkys \and
  Antanas~Vaitkus \and
  Algirdas~Grybauskas \and
  Saulius~Gražulis
}

\begin{document}

\maketitle

\section{Abstract}
X-ray crystallography rarely captures chemical bonding between atoms of a structure in question.
Most of the time distance-based heuristics are applied to establish the pairs of bonded atoms.
One class of such heuristics depends on a set of bonding radii that estimate the idealised size of each chemical element in a bond.
This publication describes an unsupervised workflow for deriving a bonding radii set from crystal structure data in the Crystallography Open Database.

\section{Introduction}
\begin{figure}
    \begin{center}
        \includegraphics[scale=.375]{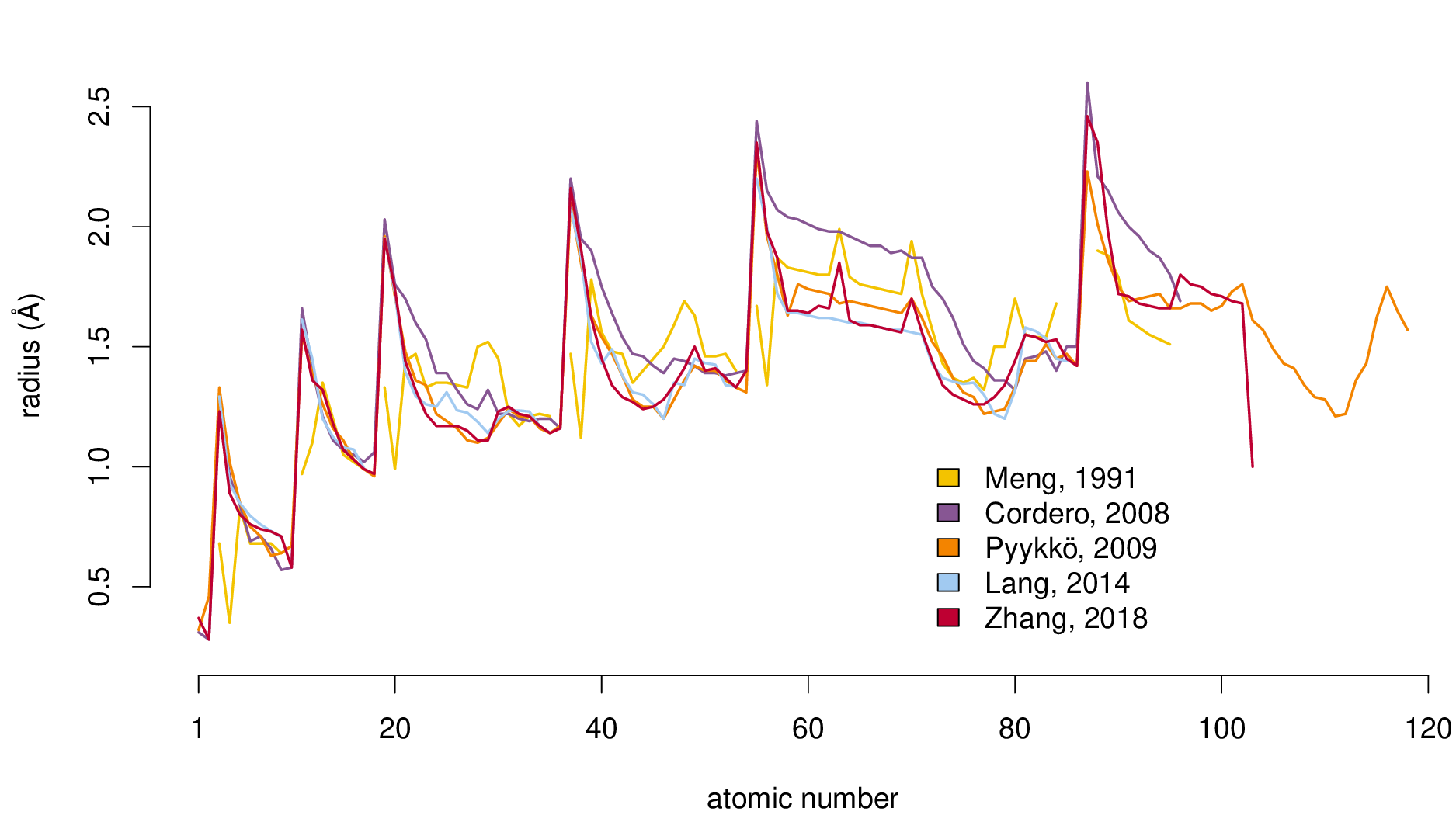}
    \end{center}
    \caption{Published bonding radii sets.}
    \label{fig:covalent-radii-published}
\end{figure}

Chemical connectivity in crystal structures is of interest to many derivative studies.
Connectivity is an unavoidable prerequisite for, to name a few, the identification of chemical substances~\cite{Merkys2023,Vaitkus2023}, analysis of molecular geometry~\cite{Merkys2018} and crystal properties~\cite{Mounet2018}.
X-ray crystallography -- a well-established technique for the determination of 3D atom coordinates in crystal structures -- is in practice rarely used to detect chemical connectivity, which, although possible using charge density analysis, is very demanding for data quality and the expertise of the researcher~\cite{Koritsanszky2001}.
Thus heuristics are often applied to derive the connectivity, usually based on per-element contribution called a \emph{bonding radius}.
This often used bond determination heuristic states that the distance between two chemically bonded atoms has to be smaller than or equal to the sum of their bonding radii~\cite{Meng1991,Baber1992,Hendlich1997,Labute2005,Zhao2007}.
Thus two atoms of chemical elements $i$ and $j$ separated by distance $d_{i,j}$ are treated as bonded if the following inequality holds:
\begin{equation}
    d_{i,j} \leq r_i + r_j + \epsilon
    \label{eqn:radii-sum}
\end{equation}
Here $r_i$ is the bonding radius of chemical element $i$ and $\epsilon$ is a sensitivity constant, usually ranging between 0.35~\AA{} and 0.45~\AA~\cite{Pan2021}.
Such radii are not constant intrinsic properties of atoms as their sizes are highly dependent on the environment~\cite{Cordero2008}.
Nevertheless bonding radii are approximated well enough and used in the distance-based connectivity criterion.

Although there are a number of published bonding radii sets, a univocal method for bonding radii derivation has not been established.
A widely used bonding radii set published as a part of the Cambridge Structural Database (CSD)~\cite{Groom2016} has received critique for manual adjustments of radii aimed to fit specific molecules as well as default values being used in place of unknown radii~\cite{Cordero2008}.
In many other studies~\cite{Meng1991,Cordero2008} bonding radii sets are derived from the CSD crystal structures that are annotated with expert-curated connectivity information.
Connectivity as derived by experienced chemists is usually of high quality, however, manual marking of connectivity is a slow and expensive process.
Moreover, the CSD operates under a proprietary license, restricting the spread of the derivative datasets and preventing the replicability of study results.
In addition, values in bonding radii sets can differ quite significantly, in some cases exhibiting differences of over 1~\AA{}, as evident from Figure~\ref{fig:covalent-radii-published}.
The need for an automated bonding radii derivation approach is addressed in recent studies published by Nikolaienko and coworkers~\cite{Nikolaienko2020,Nikolaienko2021}.
However, the main limitation of these studies is that the derived datasets involve only a few select elements thus reducing their applicability.
For a more detailed review of currently available bonding radii sets see~\cite{Sidlauskaite2023}.

Bonding radii sets are usually interaction-type-specific.
In this study we aim to determine radii encompassing all of the intramolecular interactions (covalent, ionic, metallic bonding, etc.) because such radii could be used to determine bonding in crystallographic structures.
In the study published by Alvarez~\cite{Alvarez2013} an interatomic distance distribution is used to determine the typical distance between atoms that interact mainly by the van der Waals forces.
In a typical interatomic distance distribution intramolecular and intermolecular interactions are separated by the \emph{van der Waals gap} due to a significant difference in the interaction strength.
Identification of the van der Waals gap should be sufficient to determine the longest possible interatomic distance for each pair of elements that would still allow an intramolecular bond, at the same time rejecting longer intermolecular interactions.
Throughout our study we maintain the view of interatomic distance distributions as postulated by Alvarez, on the other hand acknowledging departures from it due to the vast variability of chemical environments.

In this study we have developed an automated and unsupervised workflow for the derivation of bonding radii sets from any small-molecule crystallographic data, expressed in CIF files using the IUCr Core CIF dictionary data items.
The set derived from the data in the Crystallography Open Database (COD)~\cite{Grazulis2012} is validated and compared to other similar bonding radii sets.

\section{Methods}
\subsection{Input data}

We used all 519\,021
crystal structure records from the COD revision 295629
as an input for our study.
In the COD, crystal structures are stored as the Crystallographic Information File (CIF) format~\cite{Hall1991} files and are therefore directly suitable for the workflow described here.
While the CIF core dictionary defines data names for bond lengths (\texttt{GEOM\_BOND} category), this information mostly concerns bonded interactions and is not guaranteed to be complete or free from mistakes~\cite{Merkys2018}, thus we analysed the atomic positions only.

\subsection{Structure selection criteria}

We imposed several criteria to select only complete, experimentally measured structures and reject various biases and outliers.
We strove to use only the structures having 3D coordinates for all of their sites, including the solvent molecules.
This criterion stems from the need to correctly identify direct intramolecular and intermolecular interactions.
Due to the difficulties in addressing disordered structures, we have decided to exclude them altogether.
We employed several heuristics to exclude theoretical structures and the ones having one or more atoms which were not backed by experimental evidence.
Outlier rejection also removes structures solved under unusual conditions such as high pressure.

We have excluded the following kinds of structures from our calculations:
\begin{itemize}

    \item
        {\bf Disordered structures.}
        We exclude structures that are either explicitly marked as disordered in their CIF files or that are detected as disordered by the \progname{cif\_mark\_disorder} program from the \pkgname{cod-tools} v3.10.0 package~\cite{Vaitkus2025}.
        This is done because disordered structures are difficult to accommodate in the present analysis.
        Such structures provide either alternative sites for a set of atoms (positional disorder), alternative chemical elements occupying the same site (compositional disorder), or a combination of both.
        There is no clear solution for handling such structures as input for geometric analysis.
        One possible method would be to generate all possible combinations of disordered fragments and exclude the repetitive measurements from the stable part of the structure.
        However, alternative atom positions may affect the whole Voronoi partition rather than just the immediate environment of the disordered fragment.
        Moreover, Voronoi partition may depend on disorder in neighbouring crystal cells.
        Supercell modelling approach might offset the former problem for the price of increased computation complexity.
        Due to this complexity we decided to exclude disordered structures altogether.

    \item % filters/enabled/no-partial-occupancy
        {\bf Structures with partially occupied atom sites.}
        Partially occupied sites appear in structures with vacancy defects, crystallographically disordered structures as well as improperly modelled structures.
        All these cases are difficult to accommodate in the current statistical analysis, thus the corresponding structures are excluded.

    \item % filters/enabled/no-different-no-H-formulae
        {\bf Structures with different declared and calculated formulae.}
        The difference in heavy atom numbers between formulae usually means that the described structure is incomplete.
        
    \item
        {\bf Structures having attached hydrogen atoms.}
        When the 3D coordinates of hydrogen atoms cannot be determined, their presence near a specific heavy atom is sometimes instead indicated as the number of attached hydrogen atoms (also known as virtual hydrogen atoms) using the \texttt{\_atom\_site\_attached\_hydrogens} CIF data item.
        Since in this case the 3D coordinates of the hydrogen sites are unknown, we cannot precisely identify contacts involving them.
        While it would be possible to ignore only the interactions of the atoms having attached hydrogen atoms, we have decided to exclude such structures altogether since they comprise less than 1\% of the COD.

    \item % filters/enabled/no-dummy-sites
        {\bf Structures having dummy sites.}
        Dummy sites in CIF files do not pertain to actual atom positions, but they may signal that certain atoms are missing from the structure.
        Therefore it is safer to exclude such structures.
        There are 420
        such entries in the COD in total.

    \item % filters/enabled/no-duplicates
        {\bf Duplicate entries.}
        The COD contains a couple thousand of duplicated entries, marked as such by its maintainers.
        These entries are excluded to avoid overrepresentation of certain observations.

    \item % filters/enabled/no-non-hydrogen-calc-sites
        {\bf Structures with calculated heavy atom sites.}
        We exclude structures with explicitly marked calculated sites to retain only experimentally located atoms.

    \item % filters/enabled/no-superspace-group
        {\bf Structures solved in superspace groups.}
        Superspace group structures are difficult to accommodate in our current analysis due to interatomic distances in them being dependent on modulation waves spanning many crystal cells.

    \item % filters/enabled/no-theoretical-structures
        {\bf Theoretical structures.}
        We base our insights on experimentally determined structures, thus we filter out COD entries that were marked as theoretical by the curators of the database.

    \item % filters/enabled/no-unmodelled-solvent
        {\bf Structures with unmodelled solvent.}
        Solvent is rarely interesting in structural analysis, moreover, disordered or mobile molecules may be difficult to represent in CIF files.
        Non-modelled solvent molecules are sometimes reported using the \texttt{\_platon\_squeeze\_void\_count\_electrons} or \texttt{\_smtbx\_masks\_void\_count\_electrons} CIF data items, thus we have excluded all structures containing these data items (27189
        in total).
        However, these data items are often absent, thus we also rely on other filters (formulae mismatch, unrealistic densities, etc.)\ to identify such structures.

    \item % filters/enabled/sensible-calculated-density
        {\bf Unusual crystal densities.}
        We have excluded crystals with density lower than 0.17~$\frac{\text{g}}{\text{cm}^3}$ (density of extremely low density structures as reported in~\cite{ElKaderi2007}) or higher than 22.59~$\frac{\text{g}}{\text{cm}^3}$ (density of osmium as reported in~\cite{Arblaster2014}), both declared and calculated from formula and cell parameters.
        To ensure consistency between the declared crystal density and the calculated one, we exclude structures with the ratios of said densities outside the range of $[0.75, 1.25]$.

    \item % filters/enabled/sensible-ML-density-differences
        {\bf Structures with large absolute density differences between original and relaxed form.}
        To exclude incomplete or otherwise unrealistic crystal structures we have removed from consideration the ones which significantly diverge when relaxed using the M3GNet~\cite{Chen2022} graph deep learning interatomic potential (Justinas Šlepavičius, personal communication).
        It should be noted, however, that the relaxation appropriately preserves naturally porous structures such as metal-organic frameworks, as their framework structure is strong enough to be kept in place.
        Since the relaxation process can considerably alter both the cell parameters and atom coordinate values without actually changing the structure (this usually happens when there is a change in the unit cell basis), we based our exclusion criterion on the difference of crystal structure densities.
        Thus we exclude structures undergoing absolute density change of $0.5\frac{\text{g}}{\text{cm}^3}$ or more.

    \item % filters/enabled/sensible-R-factor
        {\bf Structures with high R factor values.}
        We require the crystallographic R factor to be known and be $\leq 0.1$, per the IUCr validation guidelines~\cite{RFACG_01}.

    \item % filters/enabled/sensible-wR-factor
        {\bf Structures with high weighted R factor values.}
        We require the weighted crystallographic R factor (wR) to be known and be $\leq 0.25$, per the IUCr validation guidelines~\cite{RFACR_01}.

    \item % filters/enabled/sensible-temperature
        {\bf Structures measured under extreme temperature.}
        Extreme temperatures (higher than 320~K) may affect the crystal geometry.

    \item % filters/enabled/sensible-pressure
        {\bf Structures measured under extreme pressure.}
        Extreme pressures (lower than 80~kPa or higher than 120~kPa) may affect crystal geometry.

\end{itemize}

\sloppypar
Most of the filters are constructed using the output produced by the \progname{COD::CIF::Data::CIF2COD} and \progname{COD::CIF::Data::CODFlags} modules from \pkgname{cod-tools}.
All filters were applied sequentially on the input data.
In total, 294\,702
entries were used for further analysis.

\subsection{Structure processing workflow}

The COD contains crystal structures described by their asymmetric units plus their space group.
This description is enough to reconstruct all atom positions in the unit cell.
We reconstruct the symmetrically equivalent positions with the \progname{cif\_fillcell} script from \pkgname{cod-tools}~\cite{Grazulis2015}.
The same tool constructs a $3 \times 3 \times 3$ supercell to represent all interactions where atoms from the central unit cell participate.

\subsection{Interaction measurements}

We have harnessed Voronoi tessellation to detect the pairs of possibly interacting atoms.
To perform the tessellation, we have chosen an algorithm implemented in \pkgname{voronota} v1.18.1877~\cite{Olechnovic2014}.
For each pair of directly contacting atoms \pkgname{voronota} calculates both the contact distance and the contact area which is later used as a weight of a contact.
In \pkgname{voronota}, which is mainly used to locate and describe contacts in protein structures, a probe is used to cut off distant pairwise interactions.
We have disabled the probe feature in our calculations by setting it to the longest diagonal of the supercell, as we are interested in all pairwise interactions regardless of their distance.
To use \pkgname{voronota} in our calculations, we have wrapped it into a tool named \progname{cif\_contacts} of the \pkgname{crystal-contacts} package~\cite{Merkys2025b}, revision 56.

Due to the abundance and relatively low precision of their site 3D coordinates, hydrogen atoms require special treatment.
Explicit hydrogen sites usually are imprecise and/or derived using geometric constraints.
Thus taking their contacts into consideration is unlikely to result in meaningful radii values.
Nevertheless, however imprecise, hydrogen sites act as natural ``barriers'' separating heavier atoms with distant interactions.
Therefore we use hydrogen atom sites in the input for Voronoi tessellation, but afterwards discard all the observed contacts involving them.

After finding and calculating all the contacts from the $3 \times 3 \times 3$ supercell with \progname{cif\_contacts}, we retain only those contacts that involve at least one atom from the asymmetric unit.
This is done by \progname{cif\_bonds\_angles} from the \pkgname{atomclasses} package~\cite{Merkys2018}.
By removing interactions between symmetric equivalents we prevent overrepresentation of contacts.

From the total pool of calculated interatomic distances we exclude observations shorter than 1~\AA{} or longer than 6~\AA{}.
Cutoff on the lower end is imposed to filter out steric clashes, possibly arising due to unmarked and undetected disorder.
Cutoff on the upper end is supposed to reduce the influence of very distant interactions, for example, in MOFs or other porous crystals.
The upper limit is derived based on the longest possible covalent bond permitted by the covalent radii set of Cordero et al.~\cite{Cordero2008} (5.2~\AA{} for Fr–Fr).

\subsection{Determination of the bond peak}

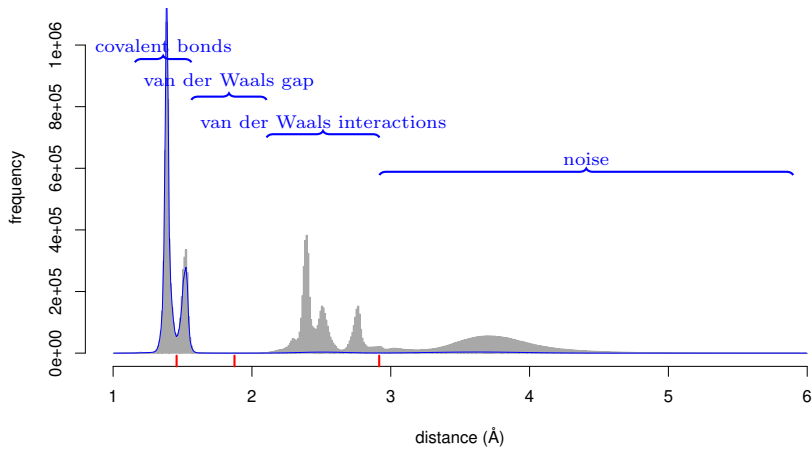
\begin{figure}
    \begin{center}
        \begin{tikzpicture}
            \node at (0,0) {\includegraphics[scale=.35]{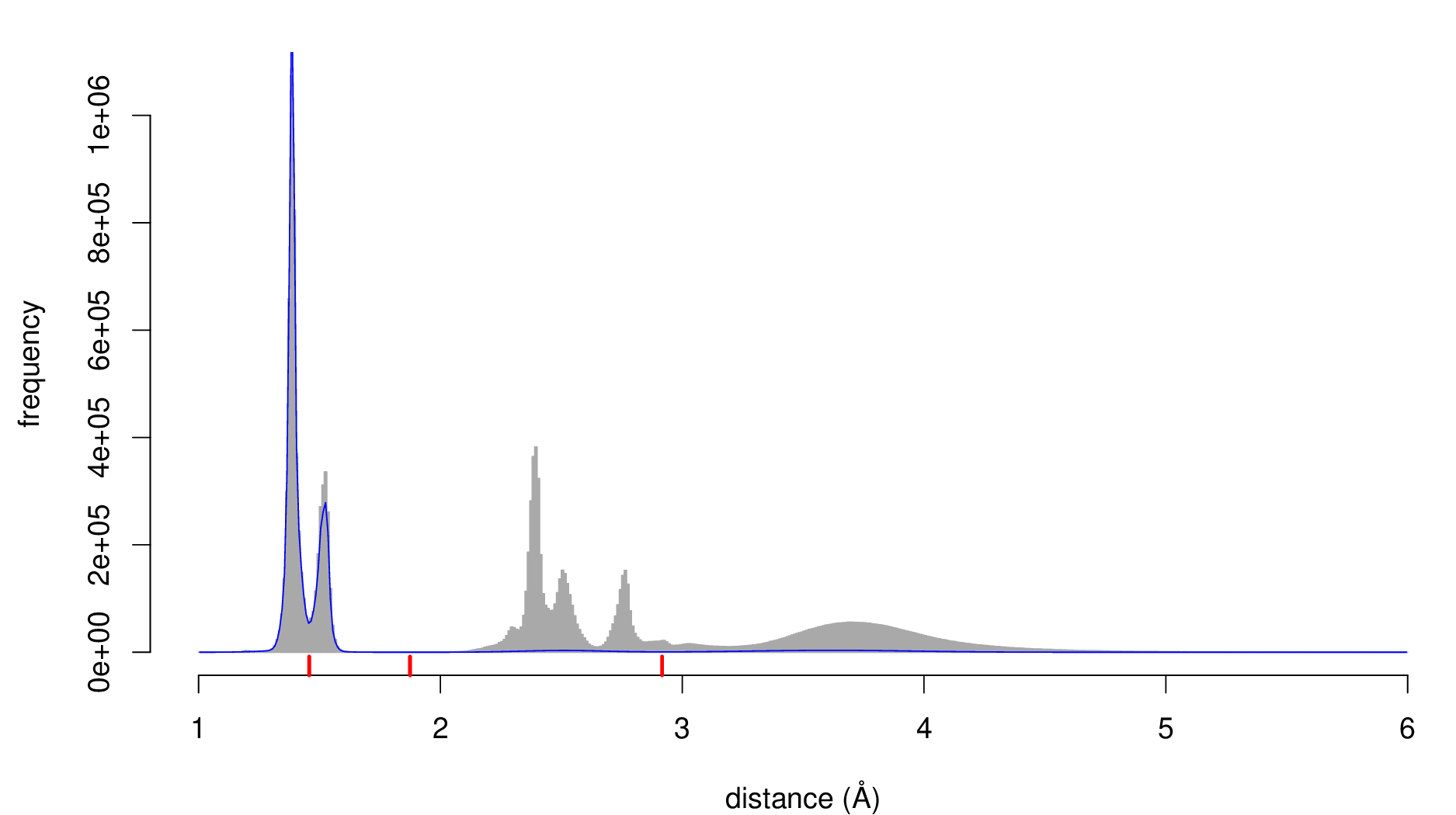}};
            \draw [thick,color=blue,decoration={brace,mirror,raise=0.5cm},decorate] (-3,1.5)
            -- node[above = 0.5cm] {\scriptsize covalent bonds} (-3.75,1.5);
            \draw [thick,color=blue,decoration={brace,mirror,raise=0.5cm},decorate] (-2,1)
            -- node[above = 0.5cm] {\scriptsize van der Waals gap} (-3,1);
            \draw [thick,color=blue,decoration={brace,mirror,raise=0.5cm},decorate] (-0.5,0.5)
            -- node[above = 0.5cm] {\scriptsize van der Waals interactions} (-2,0.5);
            \draw [thick,color=blue,decoration={brace,mirror,raise=0.5cm},decorate] (5,0)
            -- node[above = 0.5cm] {\scriptsize noise} (-0.5,0);

        \end{tikzpicture}
    \end{center}
    \caption{
        Histogram of C--C interatomic distances in the COD.
        Brackets mark approximate distance regions populated by observations from certain interactions.
        Blue curve shows the fitted mixture model density.
        Red ticks mark the antimodes as detected by our approach.
        Note the proportion differences between components matching covalent bonds and noncovalent interactions.
    }
    \label{fig:C-C-histogram}
\end{figure}

Having processed all pairs of directly interacting atoms in the input data, we group them by chemical elements and for each pair of elements (we call it a \emph{class}) identify the \emph{bond peak}, which we deem to contain the longest intramolecular interactions.
To facilitate the analysis of the density of a class, we fit a Gaussian mixture model on the sample composed of the class distances (Figure~\ref{fig:C-C-histogram}).
Fitting is done using the expectation maximisation algorithm with a modification to accept weights.
In order to reduce the influence of long nonbonded interactions we use the following weighting scheme for interaction observed between atoms $i$ and $j$:
\begin{equation}
    w_{i,j} = 1 - (1 - S_{i,j}/S_i)(1 - S_{i,j}/S_j)
    \label{eqn:interaction-weighting-scheme}
\end{equation}
Here $S_i$ is the total sum of all the contact areas for atom $i$, and $S_{i,j}$ is the contact area between atoms $i$ and $j$.
To find the best fitting Gaussian mixture model for each class, we use the expectation maximisation method to determine parameters for mixtures with the number of normal components ranging from 1 to 10.
Classes larger than 200\,000 observations are reduced by randomly picking 200\,000 observations from them.
On the other end, we require a minimal 1:30 ratio of model parameters to observations.
This means that for a mixture of $N$ Gaussian components to be considered, a sample needs at least $30 \times (3N-1)$~observations.
Component centres for the initial steps are taken from sample quantiles.
From these ten mixtures we select the one with an optimal Bayesian information criterion (BIC~\cite{Schwarz1978}) value.
Both the fitting and the BIC computation is performed using \pkgname{MixtureFitting}~v0.5.0~\cite{Merkys2025} package written in \pkgname{R}.

% From ^/trunk/radii/makefiles/Makelocal-halogens, r2549:
We have tailored the method for bond peak selection based on the chemical elements participating in the class:
\begin{itemize}
    \item For alkali and alkaline earth metals, the first mixture component is deemed to be the bond peak, as they very rarely form higher order bonds.
    \item For interactions between halogens and non-organogenic elements (not C, N, O, P, S, Se), the first mixture component is deemed to be the bond peak.
    \item For the rest, the component preceding the lowest-density antimode (supposedly corresponding to the van der Waals gap) is deemed to be the bond peak.
\end{itemize}
Different treatment of classes based on the types of elements participating in them is necessary to accommodate the variability of chemical interactions.

To select the lowest-density antimode, we inspect the density zones between each pair of neighbouring component centres.
As a midpoint between the component centers might fall on a slope of the density function, we employ the Nelder--Mead method~\cite{Nelder1965} to locate the actual minimum.
If the Nelder--Mead method locates the density minimum outside the range limited by neighbouring component centres, it is held that these components are inseparable.

\begin{figure}
    \begin{center}
        \begin{tikzpicture}
            \node at (0,0) {\includegraphics[scale=.35]{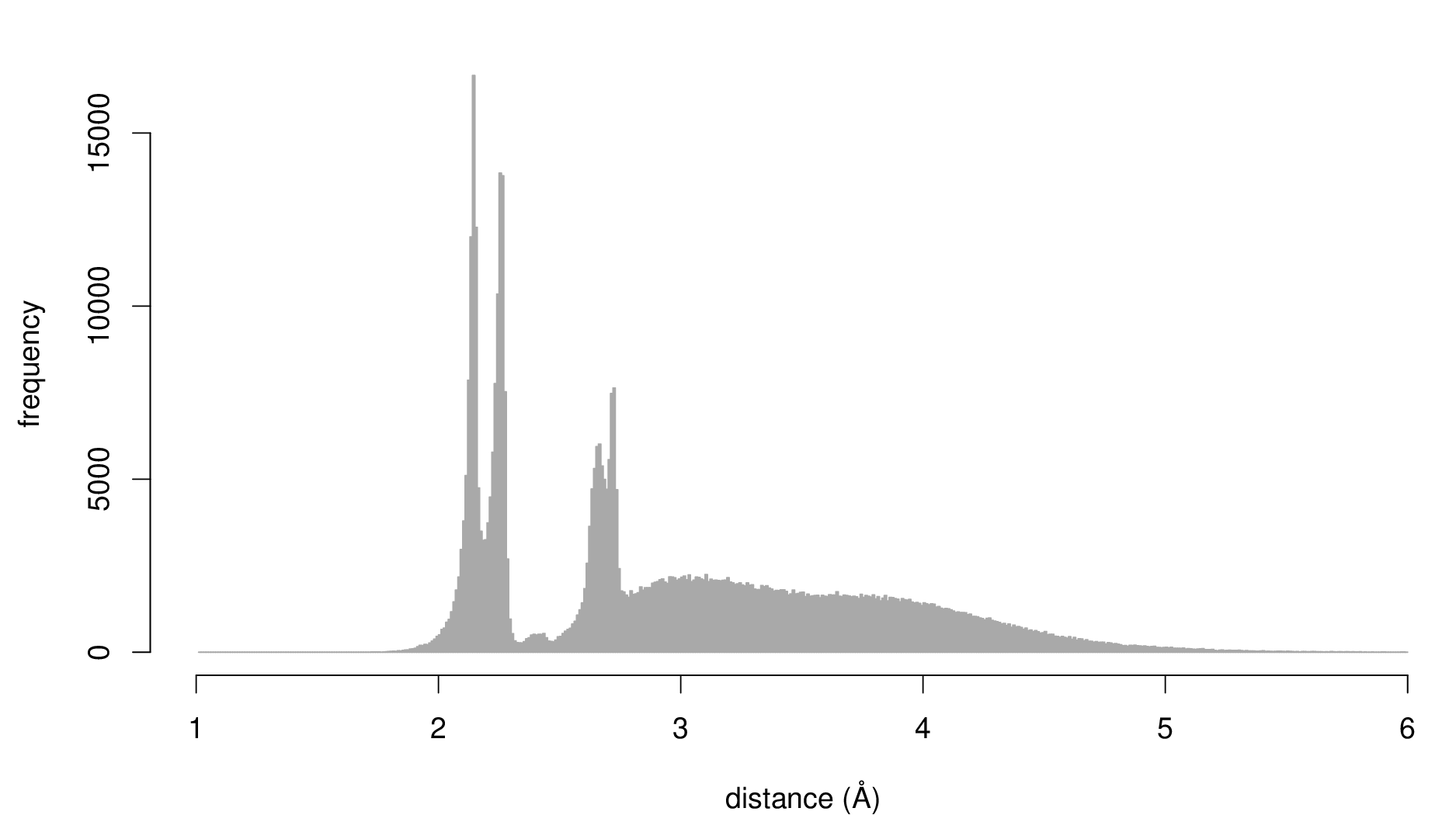}};
            \draw [thick,color=blue,decoration={brace,mirror,raise=0.5cm},decorate] (-1.85,1.5)
            -- node[above = 0.5cm] {\scriptsize --CF$_3$} (-2.05,1.5);
            \draw [thick,color=blue,decoration={brace,mirror,raise=0.5cm},decorate] (-1.65,1)
            -- node[above = 0.5cm] {\scriptsize BF$_4$} (-1.85,1);
            \draw [thick,color=blue,decoration={brace,mirror,raise=0.5cm},decorate] (3,0.5)
            -- node[above = 0.5cm] {\scriptsize $M$F$_6$ ($M$ -- metal element)} (-2.05,0.5);
            \draw [thick,color=blue,decoration={brace,mirror,raise=0.5cm},decorate] (-0.75,0)
            -- node[above = 0.5cm] {\scriptsize perfluoroaromatic compounds} (-0.95,0);
            \draw [thick,color=blue,decoration={brace,mirror,raise=0.5cm},decorate] (-1.2,-0.5)
            -- node[above = 0.5cm] {\scriptsize perfluoroalkanes} (-1.6,-0.5);
        \end{tikzpicture}
    \end{center}
    \caption{
        Histogram of F--F interatomic distances in the COD.
        Brackets mark approximate distance regions populated by observations from certain chemical compounds.
    }
    \label{fig:F-F-histogram}
\end{figure}
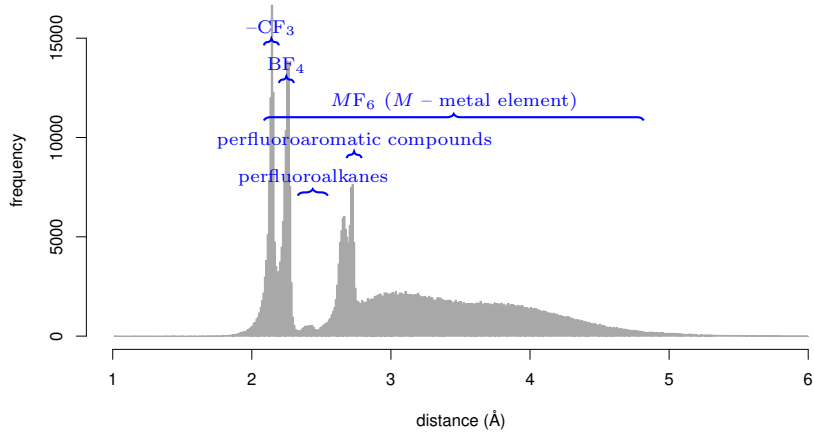

More rules dictated by the common chemical sense could be further applied to select the correct bond peaks.
In one instance, the F--F class, we have decided to completely remove the class from consideration due to its negative influence on the overall bonding radii derivation (described in Section~\ref{sec:bonding-radii}).
Proper F--F bonds are too rare in the COD to form a detectable bond peak (Figure~\ref{fig:F-F-histogram}), thus a peak formed by non-bonded interaction distances is chosen as the bond peak.
No van der Waals gap is visible.
While such aberrations are possible under our methodology, we expect them to be dominated by correctly determined bond peaks.
Unfortunately, F--F class is the most populated among all F-involving classes, thus the overly long bonding distance is assigned the largest weight.
In order to offset the influence of F--F class we have decided to remove this class altogether.

\subsection{Bonding radii}
\label{sec:bonding-radii}

We have derived the bonding radii by solving an overdetermined equation system via the least squares method:
\begin{equation}
    \begin{cases}
    W_\text{Ac,Ac} (r_\text{Ac} + r_\text{Ac}) = W_\text{Ac,Ac} d_\text{Ac,Ac} \\
    W_\text{Ac,Ag} (r_\text{Ac} + r_\text{Ag}) = W_\text{Ac,Ag} d_\text{Ac,Ag} \\
    W_\text{Ac,Al} (r_\text{Ac} + r_\text{Al}) = W_\text{Ac,Al} d_\text{Ac,Al} \\
    ... \\
    W_\text{Zr,Zr} (r_\text{Zr} + r_\text{Zr}) = W_\text{Zr,Zr} d_\text{Zr,Zr}
    \label{eq:least-squares}
    \end{cases}
\end{equation}
where $r_{i}$ is the bonding radii for chemical element $i$, $d_{i,j}$ is the bond peak centre in class of chemical elements $i$ and $j$ (including $i = j$), and $W_{i,j}$ is the weight of the equation equal to the number of observations in the class.
Weights are used to increase the influence of larger classes in the least squares calculation, as their models should be more accurate.
\pkgname{R} package \pkgname{Rlinsolve}~v0.3.2~\cite{You2021} was used to solve the system of equations.

Since both the mixture model fitting and the antimode selection could be affected by peculiarities of certain classes, we have applied an iterative outlier removal procedure to arrive to self-consistent radii set.
During each iteration, Equation~\ref{eq:least-squares} is solved and the largest class having the difference between its bond peak and radii sum greater than 0.35~\AA{} is selected.
Then the selected class is excluded from Equation~\ref{eq:least-squares} and the next iteration commences.
This is repeated until no classes with $d_\text{i,j} - (r_\text{i} + r_\text{j}) > 0.35$~\AA{} are found.
It is important to note that the difference in this criterion is not absolute, that is, only classes with bond peak position larger than radii sum are excluded.
This is done in order to eliminate classes where the bond peak centres might be detected incorrectly due to the relative abundance of long intermolecular distances.

We have also attempted to use the iterative reweighting approach for the least squares calculation (IRLS)~\cite{Wolke1988}, which iteratively decreases the weights of equations producing the largest errors in least squares calculation.
However, IRLS did not perform better than our iterative outlier removal approach, which is specifically tailored to alleviate the shortcomings of our methods.

\section{Results}
\subsection{Overview}

\begin{figure}
    \begin{center}
        \begin{subfigure}{0.45\textwidth}
            \includegraphics[scale=.325]{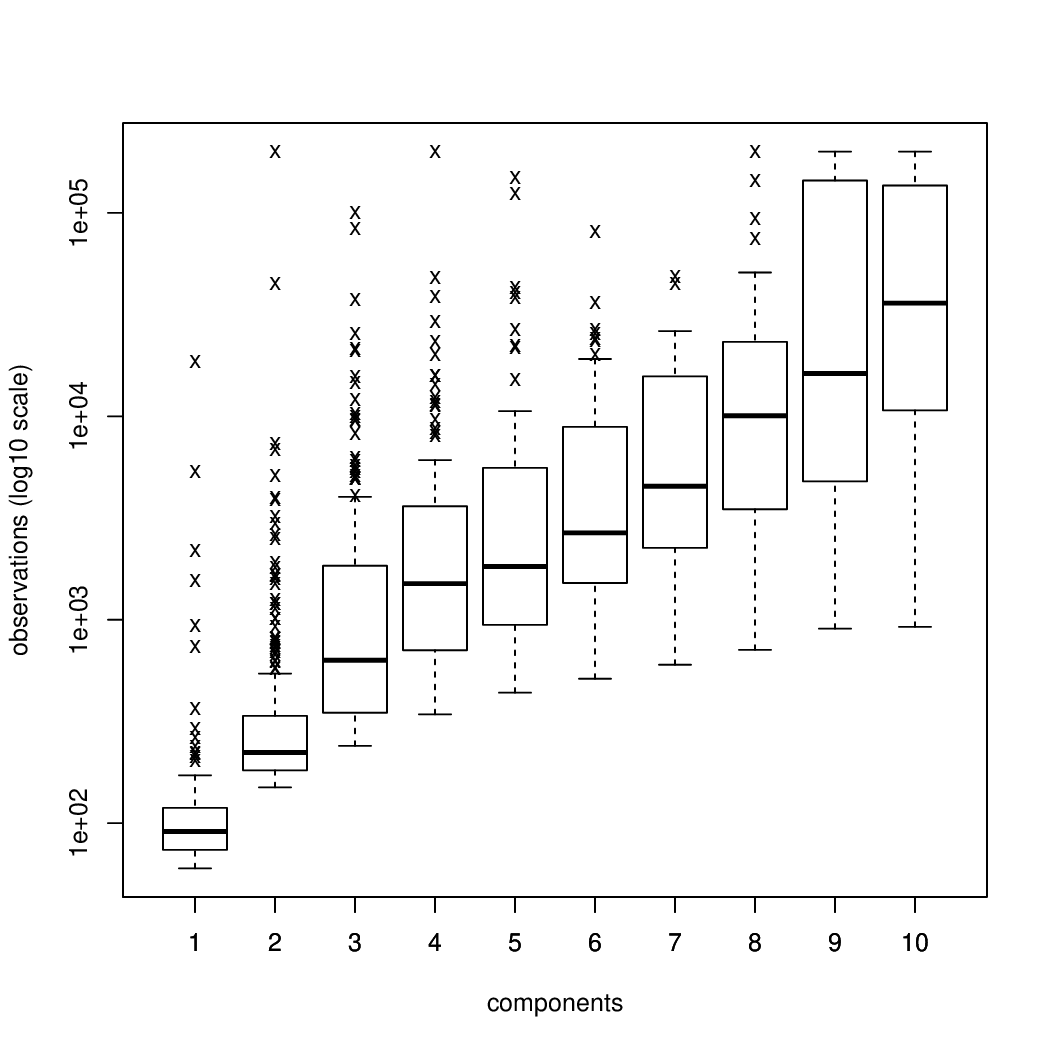}
            \caption{}
            \label{fig:components-vs-observations}
        \end{subfigure}
        \begin{subfigure}{0.45\textwidth}
            \includegraphics[scale=.325]{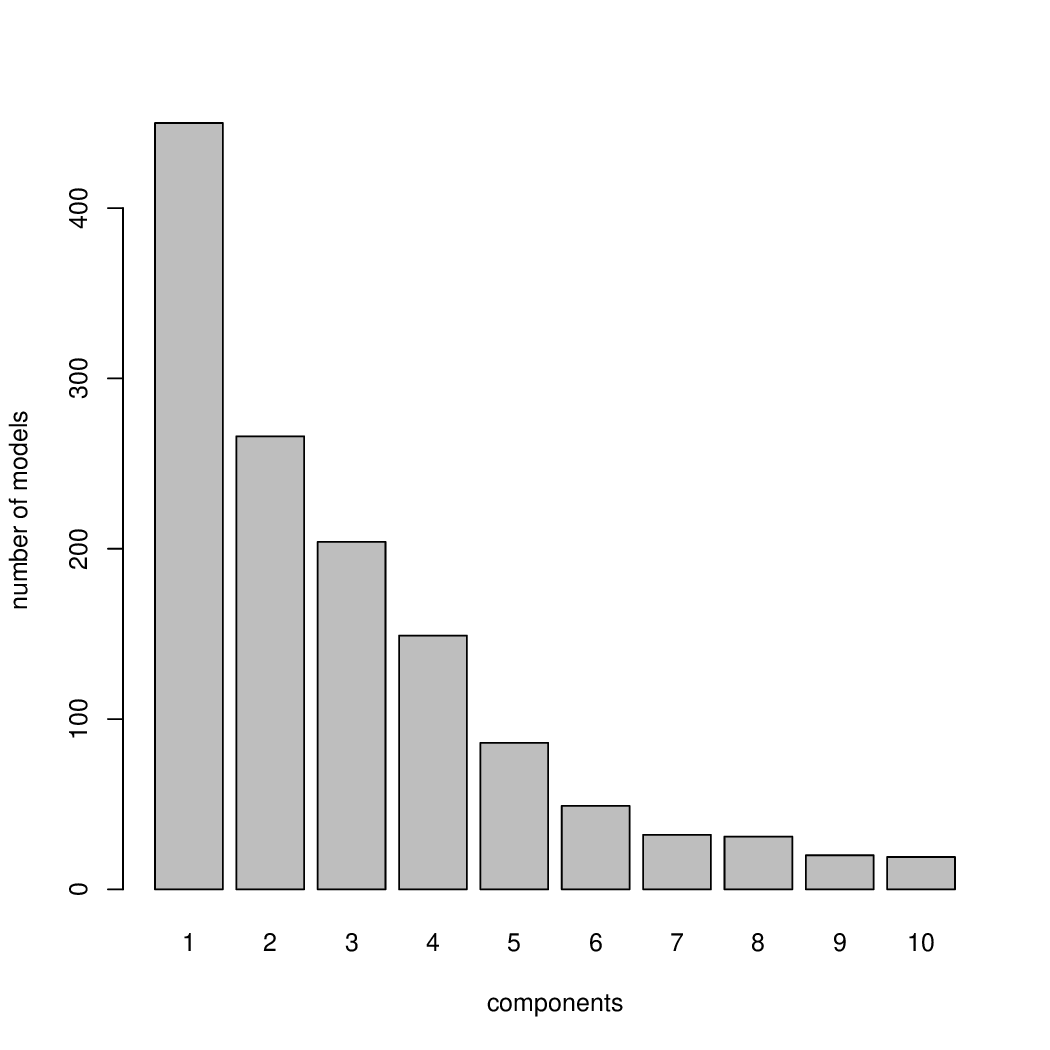}
            \caption{}
            \label{fig:components-vs-number-of-models}
        \end{subfigure}
    \end{center}
    \caption{
        Statistics of the selected models.
        (\textit{\subref{fig:components-vs-observations}}) shows the dependence of the number of mixture components in models on the number of observations, vertical axis is in $\log_{10}$ scale.
        (\textit{\subref{fig:components-vs-number-of-models}}) shows number of model components versus the number of models.
    }
    \label{fig:model-statistics}
\end{figure}

\begin{figure}
    \begin{center}
        \includegraphics[scale=.5]{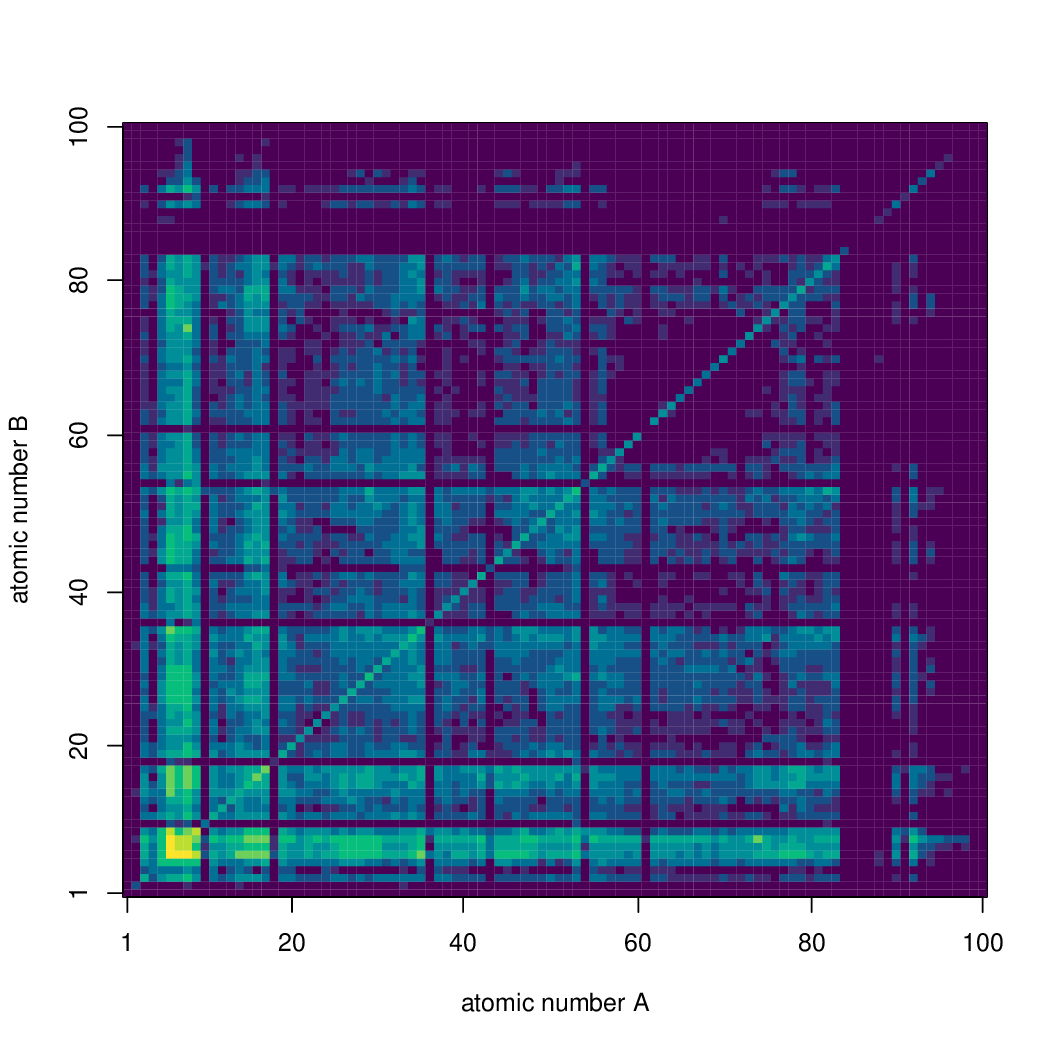}
    \end{center}
    \caption{
        Prevalence of pairwise distances between specific elements observed in the COD.
        Counts are given in $\log_{10}$ scale, with dark blue corresponding to no observations and bright yellow to most abundant.
    }
    \label{fig:class-abundancy}
\end{figure}

\begin{table}
    \scriptsize
    \begin{subtable}[t]{0.3\textwidth}
\begin{tabular}[t]{l r r}
\hline
atom & radius (\AA) & Z \\
\hline
H & - & 1 \\
He & - & 2 \\
Li & 1.28 & 3 \\
Be & 1.02 & 4 \\
B & 0.88 & 5 \\
C & 0.76 & 6 \\
N & 0.68 & 7 \\
O & 0.69 & 8 \\
F & 0.59 & 9 \\
Ne & - & 10 \\
Na & 1.70 & 11 \\
Mg & 1.42 & 12 \\
Al & 1.27 & 13 \\
Si & 1.09 & 14 \\
P & 1.02 & 15 \\
S & 0.98 & 16 \\
Cl & 0.95 & 17 \\
Ar & - & 18 \\
K & 2.01 & 19 \\
Ca & 1.71 & 20 \\
Sc & 1.60 & 21 \\
Ti & 1.39 & 22 \\
V & 1.49 & 23 \\
Cr & 1.33 & 24 \\
Mn & 1.41 & 25 \\
Fe & 1.30 & 26 \\
Co & 1.30 & 27 \\
Ni & 1.33 & 28 \\
Cu & 1.27 & 29 \\
Zn & 1.43 & 30 \\
Ga & 1.42 & 31 \\
Ge & 1.26 & 32 \\
\hline
\end{tabular}
\end{subtable}
\begin{subtable}[t]{0.3\textwidth}
\begin{tabular}[t]{l r r}
\hline
atom & radius (\AA) & Z \\
\hline
As & 1.21 & 33 \\
Se & 1.17 & 34 \\
Br & 1.18 & 35 \\
Kr & 2.10 & 36 \\
Rb & 2.13 & 37 \\
Sr & 1.94 & 38 \\
Y & 1.68 & 39 \\
Zr & 1.76 & 40 \\
Nb & 1.34 & 41 \\
Mo & 1.52 & 42 \\
Tc & 1.38 & 43 \\
Ru & 1.37 & 44 \\
Rh & 1.34 & 45 \\
Pd & 1.32 & 46 \\
Ag & 1.58 & 47 \\
Cd & 1.59 & 48 \\
In & 1.47 & 49 \\
Sn & 1.45 & 50 \\
Sb & 1.40 & 51 \\
Te & 1.45 & 52 \\
I & 1.38 & 53 \\
Xe & 1.25 & 54 \\
Cs & 2.31 & 55 \\
Ba & 2.05 & 56 \\
La & 1.85 & 57 \\
Ce & 1.83 & 58 \\
Pr & 1.84 & 59 \\
Nd & 1.84 & 60 \\
Pm & - & 61 \\
Sm & 1.85 & 62 \\
Eu & 1.74 & 63 \\
Gd & 1.73 & 64 \\
\hline
\end{tabular}
\end{subtable}
\begin{subtable}[t]{0.3\textwidth}
\begin{tabular}[t]{l r r}
\hline
atom & radius (\AA) & Z \\
\hline
Tb & 1.71 & 65 \\
Dy & 1.58 & 66 \\
Ho & 1.70 & 67 \\
Er & 1.70 & 68 \\
Tm & 1.70 & 69 \\
Yb & 1.75 & 70 \\
Lu & 1.63 & 71 \\
Hf & 1.62 & 72 \\
Ta & 1.45 & 73 \\
W & 0.64 & 74 \\
Re & 1.34 & 75 \\
Os & 1.41 & 76 \\
Ir & 1.39 & 77 \\
Pt & 1.30 & 78 \\
Au & 1.31 & 79 \\
Hg & 1.37 & 80 \\
Tl & 1.52 & 81 \\
Pb & 1.76 & 82 \\
Bi & 1.60 & 83 \\
Po & - & 84 \\
At & - & 85 \\
Rn & - & 86 \\
Fr & - & 87 \\
Ra & - & 88 \\
Ac & - & 89 \\
Th & 1.77 & 90 \\
Pa & - & 91 \\
U & 1.08 & 92 \\
Np & 1.15 & 93 \\
Pu & 1.04 & 94 \\
\hline
\end{tabular}
\end{subtable}

    \caption{
      Bonding radii set derived using the described approach.
    }
    \label{tab:radii-set}
\end{table}

In total, 3\,007
contact classes were observed, consisting of 95
unique chemical elements.
Mixture models were derived for 1\,306
classes.
This number excludes classes with too few observations to reliably fit even a one-component Gaussian mixture model.
Correspondingly, coverage of the periodic table decreased to 84
chemical elements.
Although BIC was applied as a measure to prefer simpler mixture models, there is a tendency of larger classes to be approximated with mixtures containing greater number of components, as seen in Figure~\ref{fig:model-statistics}.

The resulting radii set (Table~\ref{tab:radii-set}) is close to the one derived by Pyykk\"o and Atsumi~\cite{Pyykko2009}, especially in treatment of organogenic, alkali and alkaline earth metals where the differences for most elements do not exceed $0.05$~\AA{}.
Notable departures in these sections of the periodic table are the radii of F (discussed in Section~\ref{sec:applicability}) and Na, both significantly longer than the ones by Pyykk\"o and Atsumi.
The largest deviations between these two radii sets are observed in the d-block as well as trans-uranium elements.

\subsection{Validation}

\subsubsection{Comparison with a chemical perception pipeline}
\label{sec:validation-cif-perceive-chemistry}

\begin{figure}
    \centering
    \begin{subfigure}{0.3\linewidth}
        \includegraphics[width=\linewidth]{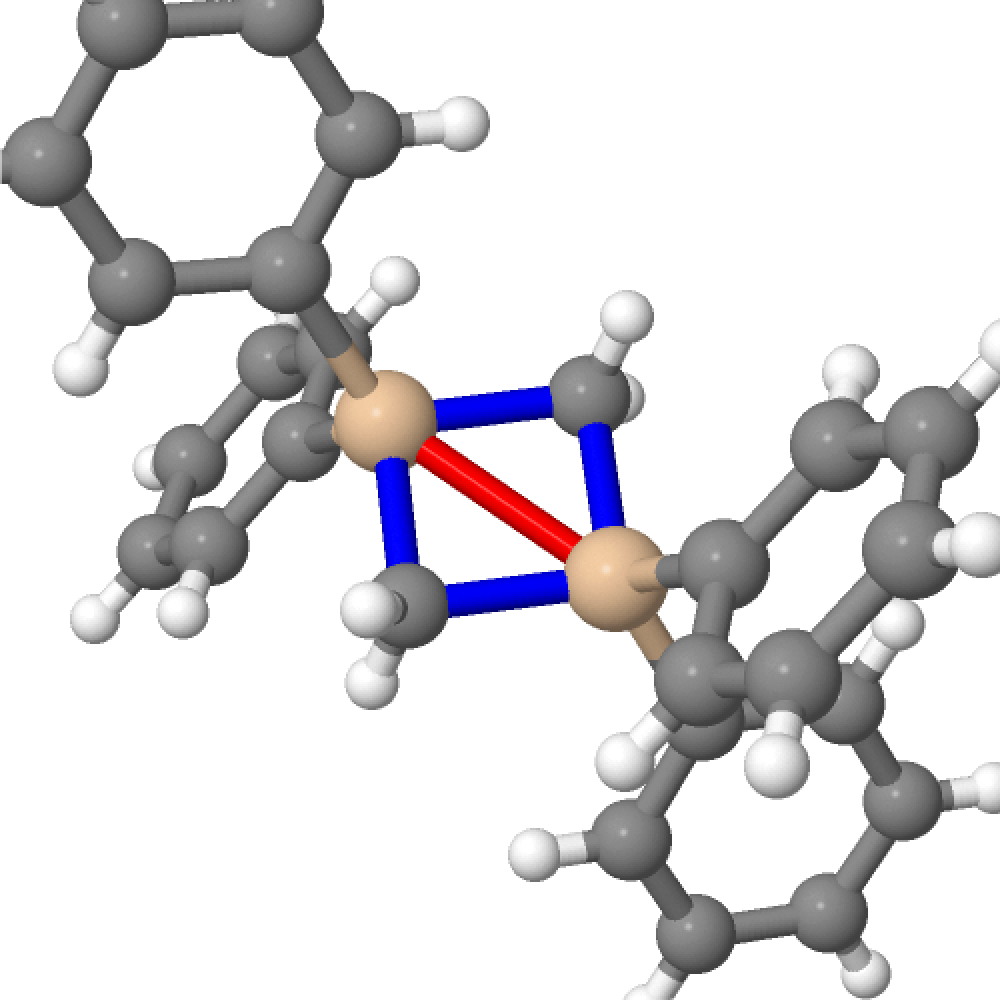}
        \caption{}
        \label{fig:connectivity-4-membered-rings-a}
    \end{subfigure}
    \begin{subfigure}{0.3\linewidth}
        \includegraphics[width=\linewidth]{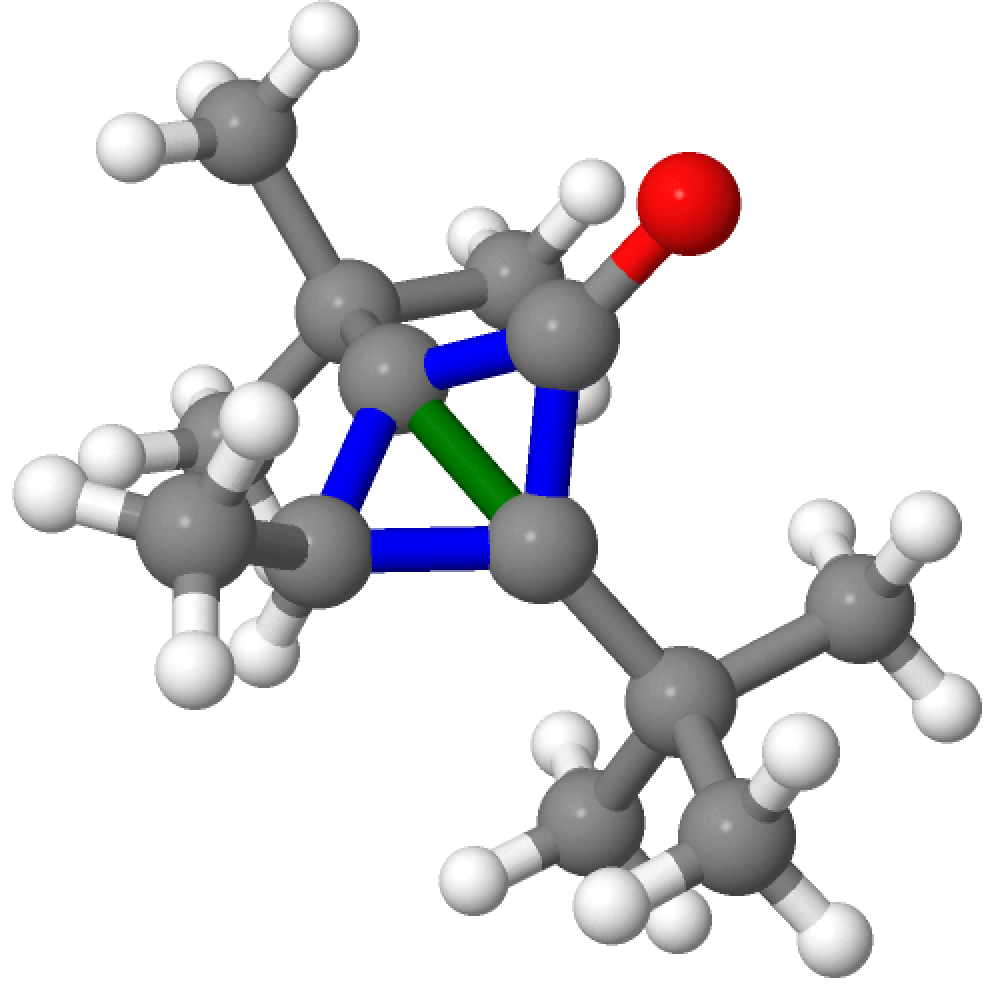}
        \caption{}
        \label{fig:connectivity-4-membered-rings-b}
    \end{subfigure}
    \begin{subfigure}{0.3\linewidth}
        \includegraphics[width=\linewidth]{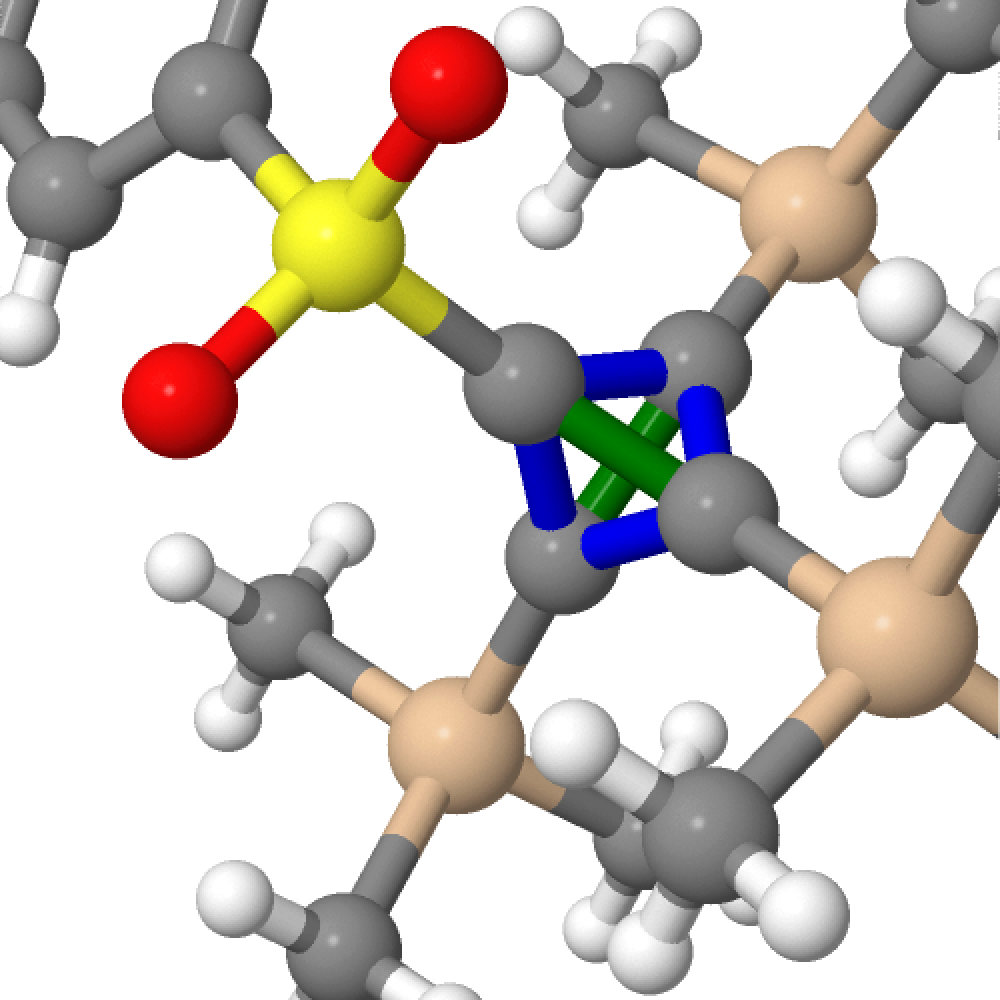}
        \caption{}
        \label{fig:connectivity-4-membered-rings-c}
    \end{subfigure}
    \caption{
        Examples of intraring bonds in 4-membered rings that would be detected in COD entries \codid{4077811} (\textit{\subref{fig:connectivity-4-membered-rings-a}}), \codid{4100665} (\textit{\subref{fig:connectivity-4-membered-rings-b}}) and \codid{4106489} (\textit{\subref{fig:connectivity-4-membered-rings-c}}) when judged on the atom radii sum criterion alone.
        Ring connections are coloured blue, legitimate intraring connections are coloured green, and false-positive intraring bonds are coloured red.
        While the diagonal intraring C–C bonds in bicyclobutane (\textit{\subref{fig:connectivity-4-membered-rings-b}}) and tetrahedrane (\textit{\subref{fig:connectivity-4-membered-rings-c}}) derivatives are legitimate, the Si–Si bond in (\textit{\subref{fig:connectivity-4-membered-rings-a}}) is a false-positive.
        Given the angle sums of 360\textdegree{} (\textit{\subref{fig:connectivity-4-membered-rings-a}}), 336\textdegree{} (\textit{\subref{fig:connectivity-4-membered-rings-b}}) and 237\textdegree{} (\textit{\subref{fig:connectivity-4-membered-rings-c}}) and the [340, 380]\textdegree{} ring flatness restriction, only the incorrect Si–Si bond would be removed during the bond pruning stage.
    }
    \label{fig:connectivity-4-membered-rings}
\end{figure}

\begin{table}[t]
    \centering
    \begin{tabular}{l r r r}
\hline
Class & New bonds & Missing bonds & Common bonds \\
\hline
Mo--Mo	&	5290	&	0	&	1196	\\
V--V	&	3520	&	0	&	6	\\
Ag--Ag	&	2994	&	0	&	2199	\\
Cu--Cu	&	2989	&	0	&	2114	\\
Fe--Fe	&	2790	&	0	&	206	\\
\multicolumn{4}{l}{...}	\\
C--C	&	52	&	0	&	7534316	\\
C--O	&	24	&	0	&	956486	\\
C--S	&	0	&	0	&	168082	\\
C--P	&	0	&	1	&	201894	\\
C--N	&	0	&	2	&	1463763	\\
\multicolumn{4}{l}{...}	\\
C--U	&	0	&	3985	&	2	\\
N--U	&	0	&	4100	&	511	\\
O--U	&	0	&	4893	&	3253	\\
O--W	&	0	&	5904	&	1484	\\
C--W	&	0	&	9419	&	200	\\
\hline
\end{tabular}

    \caption{
        Results of an attempt to reproduce connectivities derived by a chemical perception pipeline~\cite{Vaitkus2023} using our bonding radii set.
        Only 5 classes with the largest number of differences are shown for each kind of difference.
        In addition, five classes involving carbon and other organogenic elements are provided for reference.
        Column ``New bonds'' lists the number of newly assigned bonds by our radii, while ``Missing bonds'' lists unreproduced bonds.
        Column ``Common bonds'' lists the number of reproduced bonds.
    }
    \label{tab:validation-summary}
\end{table}

The derived bonding radii set was evaluated for its suitability for interatomic bonding determination in crystallographic data.
For this purpose, we analysed the bonding presented in a collection of chemically annotated crystallographic structures in SDF format that were previously derived from the COD data using a chemical perception pipeline~\cite{Vaitkus2023}.
The chemical perception pipeline establishes the bonding based on the bonding radii set compiled by Pyykk\"o and Atsumi paired with several \textit{ad hoc} rules, therefore it can be viewed as a completely independent approach.
Only the structures which passed all of the data quality checks imposed by the pipeline (data consistency, chemical validity, etc.) are used in the analysis to ensure that the presented connectivity reflects proper intramolecular bonds.

We subjected the coordinates-only part of chemically annotated structures to connectivity detection using Equation~\ref{eqn:radii-sum} ($\epsilon = 0.45$~\AA{}) looking for two types of bonding differences: missing bonds (input structure lacks an explicit bond between atoms) and overlong bonds (input structure contains an explicit bond which is deemed to be too long by our new criteria).
The connectivity detection also reproduces some of the chemical environment-specific bond handling rules employed by the chemical perception pipeline.
For example, since diagonal connections in flat 4-membered rings generally do not represent legitimate chemical bonds but rather an artefact of the bonding determination approach, they are removed to avoid false positive reports of missing bonds (see Figure~\ref{fig:connectivity-4-membered-rings}).
This is done by identifying pairs of triangles sharing an edge and having sums of their angles within [340, 380]\textdegree{}.
Every such pair is treated as constituting a flat 4-membered ring and their shared bond is removed.

In over 365\,000 input structures 125\,499
mismatches were detected.
Around 67\%
of them are due to the new bonds assigned by our radii set, with the rest being missing (i.e.\ unreproduced) bonds.
Classes with the most differences of both kinds are shown in Table~\ref{tab:validation-summary}, with classes involving carbon and other organogenic elements provided for reference.

The majority of the newly detected bonds are between the same metal atoms, highlighting the fact that our radii for these elements are longer than the ones derived by Pyykk\"o and Atsumi.
For Mo and V elements, the elongation of bonding radii is influenced by metal-OH contacts in polyoxometalates shifting the van der Waals gap to the longer side.
In turn, longer Mo and V bonding radii treats as bonded the indirect contacts between metal atoms in \emph{M}O$_6$ or, more generally, \emph{MX}$_6$ coordination clusters (\emph{M}~$\in$~\{Mo,~V\}, \emph{X} – any element).

On the other hand, our method misses bonds between organogenic elements and U and W.
This is caused by significantly shorter bonding radii of W, U and trans-uranium elements.
Bonding radii of U and trans-uranium elements are derived solely from their contacts with oxygen, with other classes being either too small or excluded as aberrant per method described in Section~\ref{sec:bonding-radii}.
The compounds contributing to observations are exclusively oxycations: uranyl, neptunyl and plutonyl, with no observations longer than the single bond length in these compounds.
Therefore, the antimode between single and double bond components is assumed to be the van der Waals gap, resulting in short bonding radii.
The bonding radius of W is strongly affected by one O--W class model component being fitted to a few very short observations.

\subsubsection{\pkgname{MaterialsCoord} benchmark}

\begin{figure}
    \begin{center}
        \includegraphics[scale=.5]{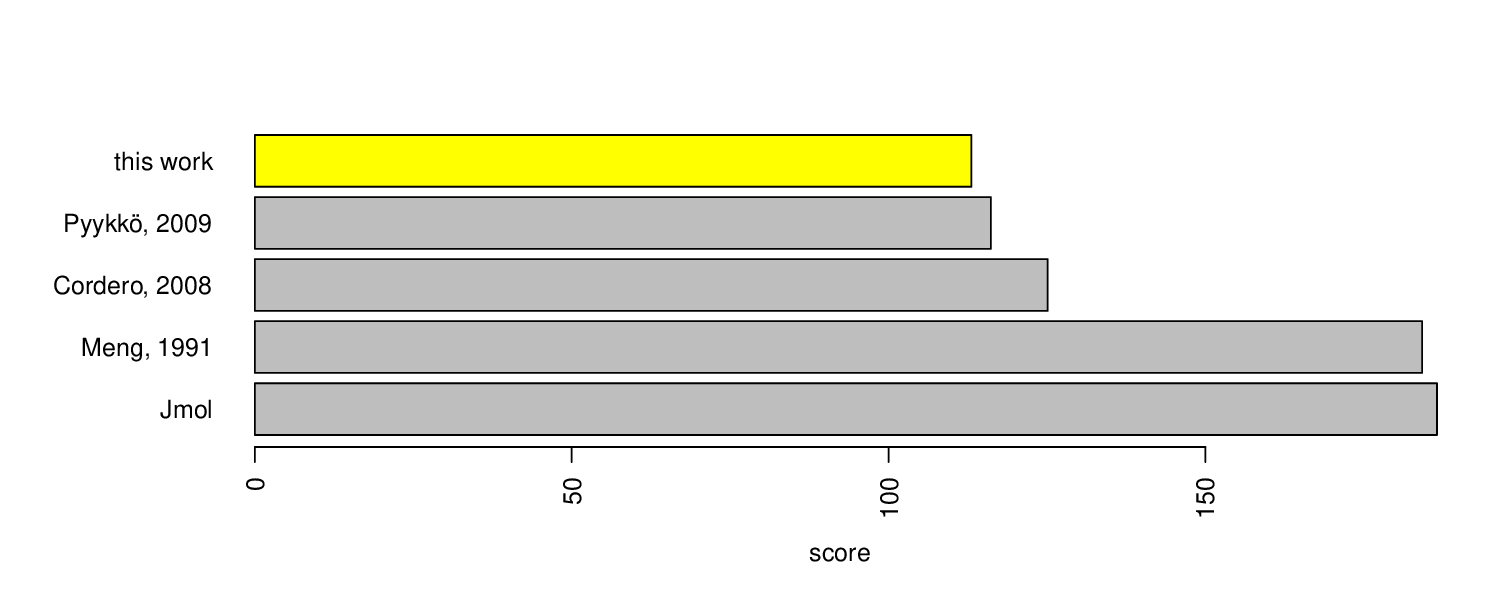}
    \end{center}
    \caption{
        Covalent radii sets validated using \pkgname{MaterialsCoord} benchmark.
        Shown are the total scores (lower is better) for the whole set of structures presented in the benchmark except the ones from \emph{clusters} group.
        A lower \pkgname{MaterialsCoord} score means a radii set yields coordination numbers closer to the expected values, with 0 being the ideal match score.
    }
    \label{fig:materialscoord-validation}
\end{figure}

We have attempted validation by plugging in our radii set to \pkgname{MaterialsCoord} benchmark~\cite{Pan2021} (Git commit \texttt{d9158f3}~\cite{Ganose2022}) which evaluates the performance of several bonding heuristics as implemented in the \pkgname{pymatgen}~v2022.11.7 package~\cite{Ong2013} for the task of predicting coordination numbers in inorganic structures.
\pkgname{MaterialsCoord} has the \texttt{JmolNN} class which implements the criterion described in Equation~\ref{eqn:radii-sum} with $\epsilon = 0.45$~\AA{}.
Due to the similarity of the \texttt{JmolNN} class to our method, we inherited \texttt{JmolNN} class to allow for easier plugging in of other radii sets, naming the derived class \texttt{CovalentRadiiTableNeighbours}.
\pkgname{MaterialsCoord} uses a set of 56 diverse experimentally determined structures, notably excluding organic compounds.
As our radii set does not define radii value for H element, we have altogether excluded the benchmark group \emph{clusters} (28 entries) containing Al$_x$H$_y$O$_z$ structures.

The comparison results are plotted in Figure~\ref{fig:materialscoord-validation}.
Although \pkgname{MaterialsCoord} does not encompass organic structures, we can see that the bonding radii set presented in this paper fares better than the earlier radii sets.
Our radii set fared consistently better than \pkgname{Jmol}'s \emph{autoBond} algorithm~\cite{Hanson2010}, the only radii sum based method described in~\cite{Pan2021}.
It should be noted that \pkgname{MaterialsCoord} does not perform diagonal bond removal from 4-membered rings, but that places our method on a par with the other compared methods.

\subsection{Radii set browser}

To better assess and compare the different algorithms and the derived sets, we have developed an interactive website~\cite{Sidlauskaite2025a}.
The website facilitates browsing of calculated radii sets, comparing between different sets and reviewing the details of radii derivation.
Radii are presented in the form of a periodic table, and a drop-down menu presents the option of switching between our radii sets and those published by other authors~\cite{Meng1991,Cordero2008,Pyykko2009}.

Clicking on a chemical element in the table opens a list of diagrams showing all the classes where the selected element participates in.
Each diagram shows a class histogram, fitted Gaussian mixture model, detected van der Waals gap and the sum of radii of the participating elements.
We have intensively used this website during the development of our methods.

\subsection{Connectivity validator}

We have created a website~\cite{Merkys2025a} exposing our connectivity validator which we have described in Section~\ref{sec:validation-cif-perceive-chemistry}.
This website accepts uploads in structural data file (SDF) and chemical markup language (CML) formats, and highlights the differences between connectivity described in the input file and the one determined using our radii set.

\section{Discussion}
\subsection{Biases}

The derived radii set is greatly skewed towards organic environments due to the abundance of organic and metalorganic compounds in the COD.
Weights of organogenic elements dominate the other elements in least squares matrix solution, some of the rarer elements have contacts with organogenic elements only.
Class sizes are shown in Figure~\ref{fig:class-abundancy}.

Classes are not free from outlier observations, warranting stricter structure selection criteria, which could result in exclusion of valid observations.
In particular, it has been observed that relaxation-based criterion does not identify structures having insignificant amounts of atoms with significant displacements from their equilibrium positions.
Moving such atoms to their equilibrium positions does not affect the unit cell, therefore no changes in crystal density are observed.
A more sensitive approach would be to identify atoms with high potential energies.

\subsection{Applicability}
\label{sec:applicability}

Our method of approximating the distribution of interatomic distances with a Gaussian mixture and then selecting an antimode as a limit between bonded and non-bonded interactions has several limitations.
All of them are related to the inability of our method to generalise over the variety of chemical interactions.

Many interaction classes do not contain pronounced gaps in their densities.
In some cases even antimodes are not present at all (as in unimodal mixtures, mixtures with components having very low proportions, or substantially overlapping components), or are not pronounced.
Cu--O class has a pair of antimodes, but both of them are still populated.
Interestingly, first of the pronounced antimodes lands in between components of equatorial and axial interactions in complexes affected by the Jahn-Teller effect, both of which we would like to detect as bonds.
Cu--Cu distance distribution is highly skewed towards longer distances and does not contain a gap; even the detected antimodes are not evident.
A special case of this problem is the unimodal distribution where it is impossible to say whether the mode contains bonded or non-bonded interactions.
Currently we treat this mode as containing bonded interactions, but this is clearly incorrect for element pairs which do not form bonds or of which bonded interactions are not observed in the COD.

Our method is sensitive to the absence and overrepresentation of certain chemical environments.
For example, F–F class is dominated by nonbonded F–F interactions, therefore identification of the van der Waals gap there is unreliable.
O–W class contains a bunch of outliers at $\approx1.3$~\AA{} which are mistaken for bond peak.
While the first case is a correct representation of F–F interactions, it needs \emph{ad hoc} handling.
The problem with O–W class has to be tackled at data level, for example by imposing stricter filtering criteria.

We have attempted to group classes by the shape of their distributions in an attempt to identify common traits and use them to apply different methods for separation of bonded and non-bonded interactions.
We have used both two-sample Kolmogorov–Smirnov test and our own method of mixture model difference integral:
\begin{equation}
    \Delta(A,B) = \displaystyle\int_0^6 | p_A(x) - p_B(x) |~\text{d}x
    \nonumber
\end{equation}
where $A$ and $B$ are classes and $p_X(x)$ is a mixture model density function for class $X$.
Both methods resulted in similar distance matrices, however, neither of them highlighted sensible clusters.
Most likely neither of the methods is sensitive enough to detect the similarities evident to the human eye without being penalised for differences, usually in location and scale.
We have attempted aligning the distributions and models using normalisation and matching of the components surrounding the perceived van der Waals gap, but none of them yielded better outcomes.

\subsection{Alternative connectivity heuristics}

While the described radii sum heuristic is very widely used in practice, there are many ongoing efforts to improve it or present alternatives to them, summarised and evaluated in, for example, Pan et al.~\cite{Pan2021}, which measures the performance of various heuristics in the task of predicting coordination numbers in crystal structures.
Some of the methods already incorporate Voronoi tessellation~\cite{OKeeffe1979,Ong2013,Zimmermann2017}.

One alternative heuristic that could be used with the results of this work is to directly rely on the element-element pairwise distances as demonstrated in~\cite{Bruno2011}.
In cases when sufficient information for a specific pair is not available, the bonding radii sum can still be calculated as the initial approximation.

\section{Conclusions}
In this publication we present a method for unsupervised derivation of sets of bonding radii from crystallographic small molecule datasets.
The devised method allows for fully automated bonding radii derivation, making the method suitable for the periodic analysis of large crystallographic resources.
This property is especially useful for deriving bonding radii from crystallographic databases, such as the COD.

In addition, we present a bonding radii set derived from the COD.
The presented set follows the trends also visible in other widely used bonding radii sets.
We have evaluated our bonding radii set by applying it on structures with known connectivities to find the apriori and derived connectivities to be in a good agreement.

It is clear that the bonding radii sum based methods will not supersede more precise methods which pay more respect to chemical environments.
However, they have their place in the computational chemist's toolbox as an indispensable approach for the derivation of connectivity when only the coordinates of atoms are known.

\section{Acknowledgements}
The authors thank Justinas Šlepavičius for allowing to use the results of his machine learning-based validation of the COD, and Kliment Olechnovič for his valuable insights into the subject matter.

\section{Funding}
This research has received funding from Vilnius University Science Support Fund under grant agreement No.~MSF-JM-15/2021 and the Research Council of Lithuania under grant agreement No.~MIP-23-87.

\section{Availability of data and materials}
The software and the data used to derive and validate the bonding radii sets described in this article are available in a Subversion repository at \texttt{svn://www.crystallography.net/contacts-in-COD}.
This article is based on revision 2657
of the repository.
All the software (scripts, libraries and workflow definitions) developed exclusively for this study are licensed under the BSD-3-Clause license, whereas all the data exclusive to this project are licensed under the CC0-1.0 license.
The repository contains embedded external software and data which are subject to their own copyright details.
To make the licensing details more explicit, we have adopted REUSE v3.3 recommendations~\cite{FSFE2024} by introducing \texttt{REUSE.toml} files throughout the repository tree.

\urlstyle{same}
\bibliographystyle{unsrturl}
\bibliography{bibliography/citations}

\end{document}